# Scalable model selection for spatial additive mixed modeling: application to crime analysis

Daisuke Murakami[1,2,*], Mami Kajita[1], Seiji Kajita[1]

[1]Singular Perturbations Co. Ltd.,
1-5-6 Risona Kudan Building, Kudanshita, Chiyoda, Tokyo, 102-0074, Japan

[2]Department of Statistical Data Science, Institute of Statistical Mathematics,
10-3 Midori-cho, Tachikawa, Tokyo, 190-8562, Japan

* Corresponding author (Email: dmuraka@ism.ac.jp)

Abstract: A rapid growth in spatial open datasets has led to a huge demand for regression approaches accommodating spatial and non-spatial effects in big data. Regression model selection is particularly important to stably estimate flexible regression models. However, conventional methods can be slow for large samples. Hence, we develop a fast and practical model-selection approach for spatial regression models, focusing on the selection of coefficient types that include constant, spatially varying, and non-spatially varying coefficients. A pre-processing approach, which replaces data matrices with small inner products through dimension reduction dramatically accelerates the computation speed of model selection. Numerical experiments show that our approach selects the model accurately and computationally efficiently, highlighting the importance of model selection in the spatial regression context. Then, the present approach is applied to open data to investigate local factors affecting crime in Japan. The results suggest that our approach is useful not only for selecting factors influencing crime risk but also for predicting crime events. This scalable model selection will be key to appropriately specifying flexible and large-scale spatial regression models in the era of big data. The developed model selection approach was implemented in the R package spmoran.

1. Introduction

Regression modeling is widely used to investigate the factors of geographical phenomena,

such as plague spread, species distribution, economic agglomeration, and crime rates. For instance, [1-3] apply regression models to study the influence of neighborhood affluence, race, unemployment rate, and facilities like liquor stores and stations, and other covariates on crime risk. Nowadays, an increasing number of open datasets of crime statistics are available [4]. For example, the Tokyo Metropolitan Government (https://www.bouhan.metro.tokyo.lg.jp/opendata/index.html), has made crime statistics (2014-present) classified by crime type and minor municipal districts publicly available. Moreover, many of the crime statistics are of good geographical resolutions at district-level or other spatial fine scales, and they record thousands to tens of thousands of data entries in each time period. For such large spatiotemporal data, a computationally efficient regression approach is very important.

In applied spatial analysis, estimation and identification of linear effects and spatially effects to objective variables are actively studied. For example, spatial econometric models are used to estimate linear effects in the presence of spatial dependence [5]. Gaussian process models have been extended to accommodate a wide variety of spatial effects in geostatistics [6]. Geographically weighted regression (GWR) [7] is used to estimate spatially varying coefficients (SVCs) on covariates [8]. Among spatial effects, we especially focus on SVC modeling that allows for estimating the local determinants of crimes [2,9,10]. For instance, [2] found that affluence reduces crime in a suburban area of Portland, Oregon, whereas it increases crime in the city center. Understanding such local differences in crime occurrence is important when considering security measures against crimes.

Apart from the spatial effects, influence from covariates can also vary non-spatially depending on time, quantile, or other variables or events [11]. Unfortunately, when all possible effects are included in the model, over-parametrization occurs and the model becomes unstable. To balance model accuracy and complexity, model selection is crucially important. For example, in crime analysis, it is needed to appropriately specify key factors behind crimes.

There are many model selection methods for SVC models [12-14] and other additive models that accommodate spatial and/or non-spatial effects [15,16]. Model selection is typically performed through iterations of model updating through inclusion/exclusion of effects (e.g., SVC) until convergence. In the case of large samples, however, SVC model selection by iterative fitting is computationally demanding. For example, mixed/semiparametric GWR [12,17], which selects constant coefficients or SVC, requires a computational time complexity of $O(N^2)$ per iteration [18], where $N$ is the sample size and $O(\cdot)$ denotes the order. Although there are fast approaches for selecting spatial and/or non-spatial effects, they still iterate model-fitting steps with a computational complexity of $O(N)$.

Given this background, this study develops a computationally efficient approach for selecting spatial and/or non-spatial effects under the framework of spatial additive mixed modeling [19,20]. It employs a pre-conditioning treatment to reduce the computation cost of parameter estimation but is not applied to model/effects selection. By extending the idea of [19,20], we develop a scalable approach for selecting both spatial/non-spatial effects. This method significantly reduces the computational time complexity of the iterative fitting steps such that the cost is independent of sample size *N*.

The remainder of this paper is organized as follows. Section 2 introduces our model. Section 3 develops our model selection procedures, and Section 4 examines its performance through Monte Carlo simulation experiments. Section 5 applies the developed approach to crime modeling and forecasting and Section 6 concludes our discussion.

2. Spatial additive mixed model

2.1. Model

We consider the following spatial additive mixed model:

$$\mathbf{y} = \sum_{p=1}^{P} \mathbf{x}_p \circ \mathbf{f}_p + \boldsymbol{\varepsilon}, \qquad \boldsymbol{\varepsilon} \sim N(\mathbf{0}, \sigma^2 \mathbf{I}), \qquad (1)$$

where $\circ$ is the element-wise product operator, $\mathbf{y}$ is a vector of response variables ($N \times 1$), where $N$ is the sample size, $\mathbf{x}_p$ is a vector of the *p*-th covariate, $\boldsymbol{\varepsilon}$ is a vector of disturbances with variance $\sigma^2$, $\mathbf{0}$ is a vector of zeros, and $\mathbf{I}$ is an identity matrix. $\mathbf{f}_p$ is a vector of coefficients describing the influence of the *p*-th covariate.[1]

There are many specifications for $\mathbf{f}_p$. The most basic specification is the constant $\mathbf{f}_p = b_p \mathbf{1}$, where $b_p$ is a coefficient and $\mathbf{1}$ is a vector of ones, which is assumed in common linear regression models.

One key idea used in this study is to specify $\mathbf{f}_p$ SVCs. For instance, [21] adopted the following specification:

$$\mathbf{f}_p = b_p \mathbf{1}_p + \frac{\tau_p^{(s)}}{\sigma} \mathbf{E}^{(s)} \boldsymbol{\Lambda}^{\alpha_p} \mathbf{u}_p^{(s)}, \qquad \mathbf{u}_p^{(s)} \sim N(\mathbf{0}_p, \sigma^2 \mathbf{I}_p). \qquad (2)$$

$\mathbf{E}^{(s)}$ is a ($N \times L_p$) matrix of $L_p$ eigenvectors corresponding to positive eigenvalues, which are called Moran eigenvectors[2]; they are extracted from a doubly-centered spatial proximity matrix [23].

---

[1] While it is more usual to express the vector of coefficient as $\mathbf{f}_p(\mathbf{z}_p)$ where $\mathbf{z}_p$ represents a vector of covariates describing the *p*-th effect, we express it as $\mathbf{f}_p$ because it is simpler and $\mathbf{z}_p$ is not necessarily needed in the subsequent discussions.

[2] One characteristic advantage of the Moran eigenvector approach is that the resulting SVC ($\mathbf{f}_p$) is interpretable through the Moran coefficient, which is a diagonal statistic of spatial dependence and its value can be positive (negative) in the presence of positive (negative) spatial dependence. Specifically, when considering all the MEs that have positive eigenvalues, $\frac{\tau_p^{(s)}}{\sigma} \mathbf{E}^{(s)} \boldsymbol{\Lambda}^{\alpha_p} \mathbf{u}_p$ describes a positively dependent map pattern. In other words, Eq. (2) furnishes a positively dependent spatial process, which is dominant in many real-world cases [22] efficiently. The Moran coefficient value increases as $\alpha_p$ grows.

$\Lambda$ is a $L_p \times L_p$ diagonal matrix whose elements are positive eigenvalues. $\mathbf{u}_p^{(s)}$ is a $(L_p \times 1)$ vector of random variables that acts as a Gaussian prior to stabilizing the SVC estimations. The $\alpha_p$ and $\tau_p^{(s)}$ parameters determine the scale and standard error of the spatial process.

Alternatively, the $\mathbf{f}_p$ function can also be determined in terms of a non-spatially varying coefficient (NVC). This specification captures the influence varying with respect to the covariate $\mathbf{x}_p$ as follows:

$$\mathbf{f}_p = b_p \mathbf{1}_p + \frac{\tau_p^{(n)}}{\sigma} \mathbf{E}_p^{(n)} \mathbf{u}_p^{(n)}, \qquad \mathbf{u}_p^{(n)} \sim N(\mathbf{0}_p, \sigma^2 \mathbf{I}_p), \tag{3}$$

where $\mathbf{E}_p^{(n)}$ is a $N \times L_p$ matrix of $L_p$, the basis function generated from $\mathbf{x}_p$. $\tau_p^{(n)}$ denotes the variance of non-spatial effects.

The following specification, which assumes both spatially and non-spatially varying coefficients (S&NVC) on all the covariates is also possible:

$$\mathbf{f}_p = b_p \mathbf{1}_p + \frac{\tau_p^{(s)}}{\sigma} \mathbf{E}^{(s)} \Lambda^{\alpha_p} \mathbf{u}_p^{(s)} + \frac{\tau_p^{(n)}}{\sigma} \mathbf{E}_p^{(n)} \mathbf{u}_p^{(n)}, \qquad \mathbf{u}_p^{(s)} \sim N(\mathbf{0}_p, \sigma^2 \mathbf{I}_p), \qquad \mathbf{u}_p^{(n)} \sim N(\mathbf{0}_p, \sigma^2 \mathbf{I}_p). \tag{4}$$

[24] showed that the S&NVC model is robust against spurious correlations, whereas naive SVC models tend to have spurious correlations [25]. Finally, Figure 1 illustrates the coefficients we will consider in this study.

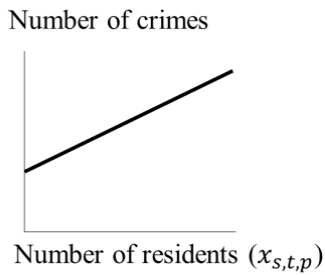
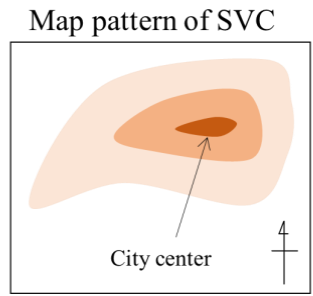
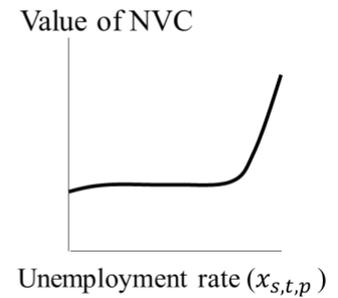

**Figure 1.** Examples of the coefficients we will consider. S&NVC is the sum of SVC and NVC. As illustrated here, SVC and NVC are defined by non-linear functions in a geographical space and a feature space, respectively.

**Table 1.** Specifications for $\{\mathbf{E}_p, \mathbf{V}_p(\boldsymbol{\theta}_p), \boldsymbol{\theta}_p\}$.

| Model Type | $\mathbf{E}_p$ | $\mathbf{V}_P(\boldsymbol{\theta}_P)$ | $\boldsymbol{\theta}_P$ |
|---|---|---|---|
| Constant | $\mathbf{0}$ | $\mathbf{0}$ | N.A. |
| SVC | $\mathbf{E}^{(s)}$ | $\frac{\tau_p^{(s)}}{\sigma}\boldsymbol{\Lambda}^{\alpha_p}$ | $\{\frac{\tau_p^{(s)}}{\sigma}, \alpha_p\}$ |
| NVC | $\mathbf{E}^{(n)}$ | $\frac{\tau_p^{(n)}}{\sigma}\mathbf{I}_p$ | $\frac{\tau_p^{(n)}}{\sigma}$ |
| S&NVC | $[\mathbf{E}^{(s)}, \mathbf{E}^{(n)}]$ | $\begin{bmatrix} \frac{\tau_p^{(s)}}{\sigma}\boldsymbol{\Lambda}^{\alpha_p} & \\ & \frac{\tau_p^{(n)}}{\sigma}\mathbf{I}_p \end{bmatrix}$ | $\{\frac{\tau_p^{(s)}}{\sigma}, \alpha_p\}$ |

In summary, $\mathbf{f}_p$, the constant is given by a fixed coefficient ($b_p$), while SVC, NVC, and S&NVC are specified by sums (linear combinations) of fixed and random effects. By substituting these values, the model Eq. (1) is formulated as follows:

$$\mathbf{y} = \mathbf{Xb} + \tilde{\mathbf{E}}(\boldsymbol{\Theta})\mathbf{U} + \boldsymbol{\varepsilon}, \quad \mathbf{U} \sim N(\mathbf{0}, \sigma^2 \mathbf{I}), \quad \boldsymbol{\varepsilon} \sim N(\mathbf{0}, \sigma^2 \mathbf{I}). \quad (5)$$

$\mathbf{X} = [\mathbf{x}_1, ..., \mathbf{x}_P]$, $\mathbf{b} = [b_1, ..., b_P]'$, $\mathbf{U} = [\mathbf{u}_1, ..., \mathbf{u}_P]'$, and $\boldsymbol{\Theta} \in \{\boldsymbol{\theta}_1, ..., \boldsymbol{\theta}_P\}$ where " ′ " represents the matrix transpose. $\tilde{\mathbf{E}}(\boldsymbol{\Theta}) = [\mathbf{x}_1 \circ \mathbf{E}_1 \mathbf{V}_1(\boldsymbol{\theta}_1), ..., \mathbf{x}_P \circ \mathbf{E}_P \mathbf{V}_P(\boldsymbol{\theta}_P)]$, where "a∘B" is an operator multiplying a column vector $\mathbf{a}$ element-wise, with each column of $\mathbf{B}$. The matrices $\mathbf{E}_p$, $\mathbf{V}_p(\boldsymbol{\theta}_p)$, and parameter $\boldsymbol{\theta}_p$ are defined in Table 1. Eq. (5) suggests that our model is formulated as a linear mixed-effects model [26] with fixed effects $\mathbf{Xb}$, random effects, $\tilde{\mathbf{E}}(\boldsymbol{\Theta})\mathbf{U}$ and $\boldsymbol{\varepsilon}$.

Although this study focuses on four specifications, other $\mathbf{f}_p$ functions have been proposed to represent linear effects, non-linear effects, group effects, and other effects, as outlined in [11]. Due to its flexibility, the additive mixed model is now used in many applied studies [27,28].

2.2. Estimation

Among the estimation algorithms for spatial additive mixed models, we focus on the fast restricted maximum likelihood (REML) estimation of [19], which is scalable for both sample size *N* and the number of effects *P*. The computational bottleneck here is an iterative evaluation of the restricted log-likelihood $loglik_R(\boldsymbol{\Theta})$ of Eq.(1) (or Eq.5) to numerically estimate the variance parameters $\boldsymbol{\Theta} \in \{\boldsymbol{\theta}_1, \cdots \boldsymbol{\theta}_P\}$ (see Appendix A). To reduce cost, [19] developed a sequential estimation of $\{\boldsymbol{\theta}_1, \cdots \boldsymbol{\theta}_P\}$ parameters by applying Eq. (6) until convergence.

$$\widehat{\boldsymbol{\theta}}_p = \underset{\boldsymbol{\theta}_p}{\mathrm{argmax}} \ loglik_R(\boldsymbol{\theta}_p | \boldsymbol{\Theta}_{-p}). \quad (6)$$

While the direct maximization of Eq. (6) still has the same computational burden, [19] developed the following procedure for the fast REML:

(I) Replace the data matrices $\{\mathbf{y}, \mathbf{X}, \mathbf{E}_1, \dots, \mathbf{E}_P\}$ whose dimensions are dependent on $N$, with their inner products whose dimensions are independent of $N$.
(II) Using the inner products, iterate the following calculations sequentially for $p \in \{1, \dots, P\}$:
  (II-1) Estimate $\widehat{\boldsymbol{\theta}}_p$ by maximizing $loglik(\boldsymbol{\theta}_p | \widehat{\boldsymbol{\Theta}}_{-p})$ with $\boldsymbol{\Theta}_{-p} \in \{\boldsymbol{\theta}_1, \dots \boldsymbol{\theta}_{p-1}, \boldsymbol{\theta}_{p+1}, \dots \boldsymbol{\theta}_P\}$.
  (II-2) Go to (III) if the likelihood value converges. Else, return to (II-1).
(III) Output the final model.

In this algorithm, the data matrices are replaced with their inner products before the iterative likelihood evaluation step. After all, the computational complexity of the iterative likelihood evaluation to find $\widehat{\boldsymbol{\theta}}_p$ in step (II-1) reduces to $O(L_p^3)$ [19]. In other words, after the pre-conditioning step (I), the computational complexity of the fast REML is highly scalable for both the sample size $N$ and the number of effects $P$. Owing to this property, the spatial additive mixed model Eq.(1) can be estimated efficiently even when $N$ and $P$ are very large.

## 3. Model selection

### 3.1. Introduction

As explained in Section 2, the coefficients in Eq. (1) can be constant, SVC, NVC, or S&NVC. The complexity of the model depends considerably on the selection of the coefficient type. For instance, Eq. (1) has only $P$ coefficients if all the coefficients are assumed to be constant. In the case of the SVCs-based model, on the other hand, the number of coefficient parameters is $\sum_{p=1}^{P} L_p$, since the model utilizes $L_p$-dimension Moran eigenvectors for its $P$ coefficient representations. Too many parameters can lead to overfitting and overestimation of statistical significance, whereas too few parameters can cause underfitting. This issue can be remedied by choosing an optimal model that provides an appropriate balance of the sizes of parameters and datasets with respect to model accuracy.

There is one difficulty in the selection of a large number of candidate models; there are $4^P$ model specifications for Eq. (1). For example, if $P = 9$, which we will assume later, there are $4^9 = 262,144$ models. However, in practice, it is desirable to find or approximate the best model within seconds or minutes. Hence, this study develops a computationally efficient model selection approach. We attempt to search the model by minimizing the cost function, which can be defined by the Akaike information criterion (AIC) or Bayesian information criterion (BIC).

It is important to note that constant, SVC, NVC, and S&NVC share the same fixed effect $(b_p \mathbf{1})$, whereas their random effects differ from each other. In other words, we select random effects. In such a case, REML-based AIC and BIC are available for the model selection of linear additive mixed models [29]. While there are marginal and conditional AIC/BIC specifications, we focus on marginal BIC for the following reasons:

- It is the most common specification for linear mixed effects models [30], including spatial additive mixed models.
- Poor performance of conditional AIC/BIC-based model selection was reported when considering two or more random effects (see [31,32]) whereas [33] showed that conditional AIC/BIC outperforms when comparing models with/without one random effect.
- Although the marginal specification suffers from a theoretical bias, [32] showed that the influence of the bias on model selection result is quite small.

Based on a preliminary analysis, we decided to use REML-based marginal BIC, defined by $-2loglik_R(\Theta) - 2Q\log(N)$, where $Q$ is the number of fixed coefficients ($P$) and variance parameters in $\Theta$ in the crime analysis in Section 4.

### 3.2. Model selection procedures

This section proposes novel practical methods for model selection. The first incorporates a model selection into the sequential REML estimation (see Section 2.2). To reduce the chance of trapping to local optima, the second approach relies on a Monte Carlo (MC) simulation iterating the sequential REML estimation. We call the former a simple selection method, which emphasizes simplicity and practicality, and the latter MC selection method. Sections 3.2.1 and 3.2.2. explain these methods.

### 3.2.1. Simple selection method

The simple selection method consists of model selection steps in sequential REML estimation. The procedure of this method is as follows:

(a) Replace the data matrices $\{ \mathbf{y}, \mathbf{X}, \mathbf{E}_1, \ldots, \mathbf{E}_P \}$ with the inner products as processed in step (II) in Section 2.2.
(b) Perform the following calculation sequentially for each $p \in \{1, \ldots, P\}$:
  (b-1) Estimate the $p$-th SVC by maximizing $loglik(\mathbf{\theta}_{p(s)}|\widehat{\mathbf{\Theta}}_{-p(s)})$ with respect to $\mathbf{\theta}_{p(s)}$, which is a subset of $\mathbf{\theta}_p$ characterizing the SVC and $\widehat{\mathbf{\Theta}}_{-p(s)}$ represents the set of variance parameters excluding $\widehat{\mathbf{\theta}}_p$ from $\widehat{\mathbf{\Theta}}$.
  (b-2) Select the SVC if it improves the cost function value (e.g., BIC). Otherwise, replace it with a constant.
  (b-3) Estimate the $p$-th NVC by maximizing $loglik(\mathbf{\theta}_{p(n)}|\widehat{\mathbf{\Theta}}_{-p(n)})$ with respect to $\mathbf{\theta}_{p(n)}$, which is a subset of $\mathbf{\theta}_p$ characterizing the NVC and $\widehat{\mathbf{\Theta}}_{-p(s)}$ represents the set of variance parameters excluding $\widehat{\mathbf{\theta}}_{p(n)}$ from $\widehat{\mathbf{\Theta}}$.

(b-4) The NVC is selected if it improves the cost function value (e.g., BIC). Otherwise, it is replaced with a constant.

(c) Go to (d) if the cost function converges. Otherwise, go back to (b).

(d) Output the final model.

While there are similar selection approaches [34], ours is distinctive because its computational complexity for model selection is independent of the sample size, owing to step (a) that renders all the data matrices into their inner products. Due to the drastic dimension reduction, this simple method is suitable for very large samples. Specifically, the computational complexity of the iterative part in the model selection procedure equals $O(L_p^3)$, which is required to evaluate the log likelihood value in step (b-1) and (b-3). Because the computational complexity is very small and independent of the sample size $N$, our model selection procedure is extremely fast even for large samples.

One problem is the large number of combinations of pre-determined sequence $p \in \{1, ..., P\}$. For example, if $P = 9$, there are $P! = 362,880$ sequences; some of them might result in poor model selection result (i.e., local optimum). At least, unlike the maximum likelihood estimation or cross-validation, REML tends to not have local optima [35,36]. Section 4 examines if this simple approach accurately approximates/selects the true model through Monte Carlo experiments.

### 3.2.2. Monte Carlo (MC) selection method

To reduce the risk of falling into local optima, we randomly sample sequences $p \in \{1, ..., P\}$ and iterate the REML-based estimation given the sequence. Specifically, we propose the following model selection approach:

(A) Replace the data matrices $\{\mathbf{y}, \mathbf{X}, \mathbf{E}_1, ..., \mathbf{E}_P\}$ with the inner products.

(B) Iterate the following calculation $G$ times using the inner products:

  (B-1) Randomly sample the $g$-th sequence $\{1_g, ..., P_g\}$ without replacement.

  (B-2) Perform the following calculation sequentially for each $p_g \in \{1_g, ..., P_g\}$:

   (B-2a) Estimate the $p_g$-th SVC by maximizing $loglik(\boldsymbol{\theta}_{p_g(s)}|\widehat{\boldsymbol{\Theta}}_{-p_g(s)})$ where $\{\boldsymbol{\theta}_{p_g(s)}, \widehat{\boldsymbol{\Theta}}_{-p_g(s)}\}$ are defined similarly as $\{\boldsymbol{\theta}_{p(s)}, \widehat{\boldsymbol{\Theta}}_{-p(s)}\}$.

   (B-2b) Select the SVC if it improves the cost function value (e.g., BIC). Otherwise, replace it with a constant.

   (B-2c) Estimate the $p_g$-th NVC by maximizing $loglik(\boldsymbol{\theta}_{p_g(n)}|\widehat{\boldsymbol{\Theta}}_{-p_g(n)})$ where $\{\boldsymbol{\theta}_{p_g(n)}, \widehat{\boldsymbol{\Theta}}_{-p_g(n)}\}$ are defined similarly as $\{\boldsymbol{\theta}_{p(n)}, \widehat{\boldsymbol{\Theta}}_{-p(n)}\}$.

   (B-2d) The NVC is selected if it improves the cost function value (e.g., BIC). Otherwise, it is replaced with a constant.

  (B-3) Go to (B-4) if the cost function converges. Otherwise, go back to (B-2).

  (B-4) Calculate the cost function value of the selected model.

(C) Output the best model in the selected models in terms of the lowest cost function.

As with the simple selection approach, the computational cost for iterative step (B) is independent of the sample size. In addition, the iterative step is easily parallelized. Thus, this is a computationally efficient model selection procedure. Step (B) performs a MC simulation to marginalize $g$ and obtain the distribution of the cost value. We call this approach MC selection approach.

## 4. Numerical experiments

### 4.1. Computational details

Here, we examine whether this simple selection method accurately approximates the true model, which is the model with all the coefficients types (i.e., constant, SVC, NVC, or S&NVC) being defined correctly, or if MC selection method is needed to achieve accurate model selection through comparative Monte Carlo experiments. We compare with/without the effects selection model by fitting these models to the synthetic data generated from

$$\mathbf{y} = \boldsymbol{\beta}_0 + \sum_{p=1}^{P} \mathbf{x}_p b_p + \sum_{p=1}^{P} \mathbf{x}_{1,p} \boldsymbol{\beta}_{1,p} + \sum_{p=1}^{P} \mathbf{x}_{2,p} \boldsymbol{\beta}_{2,p} + \boldsymbol{\varepsilon}, \qquad \boldsymbol{\varepsilon} \sim N(\mathbf{0}, \mathbf{I})$$

$$\boldsymbol{\beta}_0 = \tilde{\mathbf{C}} \mathbf{u}_0, \qquad \mathbf{u}_0 \sim N(\mathbf{0}, \mathbf{I}),$$

$$\boldsymbol{\beta}_{1,p} = \mathbf{1} + \tilde{\mathbf{C}} \mathbf{u}_{1,p}, \qquad \mathbf{u}_{1,p} \sim N(\mathbf{0}, \tau_{1,p}^2 \mathbf{I}),$$

$$\boldsymbol{\beta}_{2,p} = \mathbf{1} + \mathbf{E}_{2,p} \mathbf{u}_{2,p}, \qquad \mathbf{u}_{2,p} \sim N(\mathbf{0}, \tau_{2,p}^2 \mathbf{I}),$$

(7)

where the covariates $\{\mathbf{x}_1, \ldots, \mathbf{x}_1\}, \{\mathbf{x}_{1,p}, \ldots, \mathbf{x}_{1,p}\}, \{\mathbf{x}_{2,p}, \ldots, \mathbf{x}_{2,p}\}$ are generated from independent standard normal distributions. The matrix $\tilde{\mathbf{C}}$ is constructed from row standardization of spatial connectivity matrix $\mathbf{C}$ whose $(i, j)$-th element equals $\exp(-d_{i,j})$, where $d_{i,j}$ is the Euclidean distance between sample sites $i$ and $j$. The sample sites were generated from two independent standard normal distributions. The SVCs $\boldsymbol{\beta}_0$ and $\boldsymbol{\beta}_{1,p}$ are defined by spatial moving average processes. $\boldsymbol{\beta}_{2,p}$ is an NVC that varies with respect to $\mathbf{x}_{2,p}$, in which $\mathbf{E}_{2,p}$ is a matrix of 10 polynomial basis functions generated from $\mathbf{x}_{2,p}$. Figure 2 illustrates coefficients obtained from these generating processes.

The main objective is to compare the coefficient estimation accuracy and computational efficiency of simple and MC-based S&NVC model selections with alternatives. For MC model selection, we assumed 30 replicates. These models were fitted 200 times while varying $P \in \{1, 2, 3\}$, and sample size $N \in \{50, 200, 1{,}000\}$.

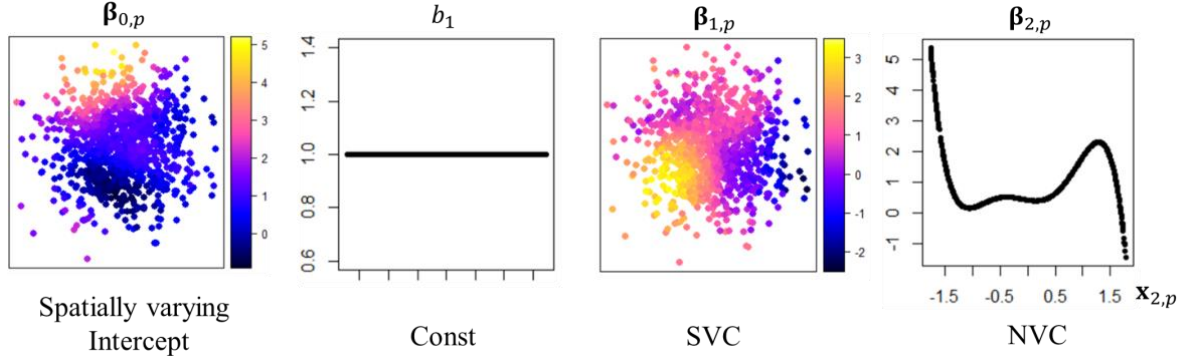

**Figure 2.** Coefficients obtained from our assumed generating processes. These patterns change in each iteration.

The coefficient estimation accuracy is evaluated using the root mean squared error (RMSE), which is defined as

$$RMSE(\boldsymbol{\beta}_{i,p}) = \sqrt{\frac{1}{200N} \sum_{iter=1}^{200} \sum_{i=1}^{N} (\hat{\beta}_{i,p}^{(iter)} - \beta_{i,p})^2}, \quad (8)$$

where *iter* represents the iteration number, $\beta_{i,p}$ is the *i*-th element of $\boldsymbol{\beta}_p$, and $\hat{\beta}_{i,p}^{(iter)}$ is the estimate given in the *iter*-th iteration. The bias of the estimates is evaluated using

$$Bias(\boldsymbol{\beta}_{i,p}) = \frac{1}{200N} \sum_{iter=1}^{200} \sum_{i=1}^{N} (\hat{\beta}_{i,p}^{(iter)} - \beta_{i,p}) \quad (9)$$

Under these settings, Section 4.2 examines if our approaches accurately select the model, while Section 4.3 compares our approaches with other approaches.

4.2. Performance of model selection

In this experiment, we compare the following six models. As baseline models, the linear regression model (LM) and the SVCs across coefficients (SVC model) are used. As another baseline, we use the model assuming true coefficients types as known (i.e., constant coefficient on $\mathbf{x}_p$, SVC on $\mathbf{x}_{1,p}$, and NVC on $\mathbf{x}_{2,p}$ (true model))[3]. We construct a full model that assumes S&NVC across coefficients (S&NVC model). In addition, we prepare simple and MC-based S&NVC model selection approaches that select their coefficients among constant, SVC, NVC, and S&NVC using the simple approach and Monte Carlo approach, respectively.

Figure 3 summarizes the RMSEs of the estimated coefficients. As expected, LM has higher estimation errors because of ignoring the spatial and non-spatial variations in regression coefficients.

---

[3] True model assumes coefficients values as unknown (i.e., only coefficients types are known), and the values are estimated from samples as same as the other models. The estimates will have an estimation error.

Although the SVC model is popular in spatial statistics, the RMSEs for the NVCs are considerably high. In addition, probably due to error, its estimation accuracy for the constant coefficients by the SVC model is worse than that of LM model. These results suggest that SVC-based models become unstable in the presence of constant or non-spatial coefficients.

Regarding the S&NVC model without model selection, the estimated SVCs and NVCs are as accurate as the true model. However, its RMSE values for the constants are the highest across the models, probably because of over-parameterization. On the other hand, the RMSEs of the simple and MC-based S&NVC model selections are close to the true model across all coefficients. These results demonstrate the importance of effect selection in spatial regression modeling.

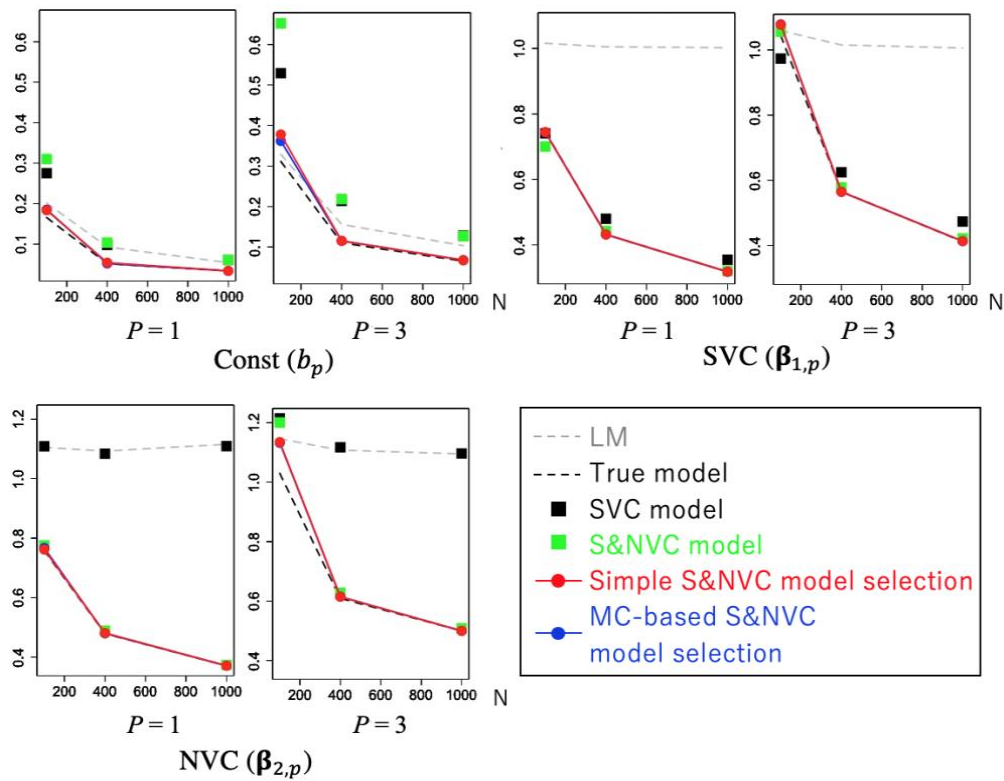

**Figure 3.** RMSEs of the coefficient estimates. $N$ denotes sample size. $P = 1$ means a model with a constant coefficient, a SVC, and an NVC in the data generating process (Eq. 7) while $P = 3$ means 3 constant coefficients, 3 SVCs, and 3 NVCs in the process.

Figure 4 plots the biases of the coefficient standard errors, which are used to evaluate statistical significance; the downward bias of the standard errors yields an overestimation of the statistical significance whereas the opposite is true for the upward bias. Here, the standard errors estimated from the true model are regarded as the true values. For standard errors of the constant coefficients estimated from LM, SVC, and S&NVC models are downwardly biased. For SVC, those estimated from LM are upwardly biased while the SVC and S&NVC models are downwardly biased. For NVC, the standard errors estimated from LM and SVC models are upwardly biased. The RMSE of the standard errors obtained by LM, SVC and S&NVC models are also high as shown in Figure 5. Based on the result, the models without effect selection (LM, SVC, and S&NVC models) suffer from an overestimation or underestimation of statistical significance.

Conversely, the standard errors estimated from the simple and MC-based model selection approaches are almost the same as those of the true model, except for the case with $P = 3$ and $N = 100$, which is the most severe case, estimating 10 effects from 100 samples. Our effects selection approaches are useful for improving estimation accuracy for both the coefficients and their standard errors. Surprisingly, it is also found that results from the simple model selection approach are almost the same as the MC-based approach, in spite of the fact that the simple approach relies on a pre-determined sequence $p \in \{1, \ldots, P\}$ for model selection, whereas the MC-based method implicitly optimizes it. This is attributable to the stability of REML (see Section 3.2.1).

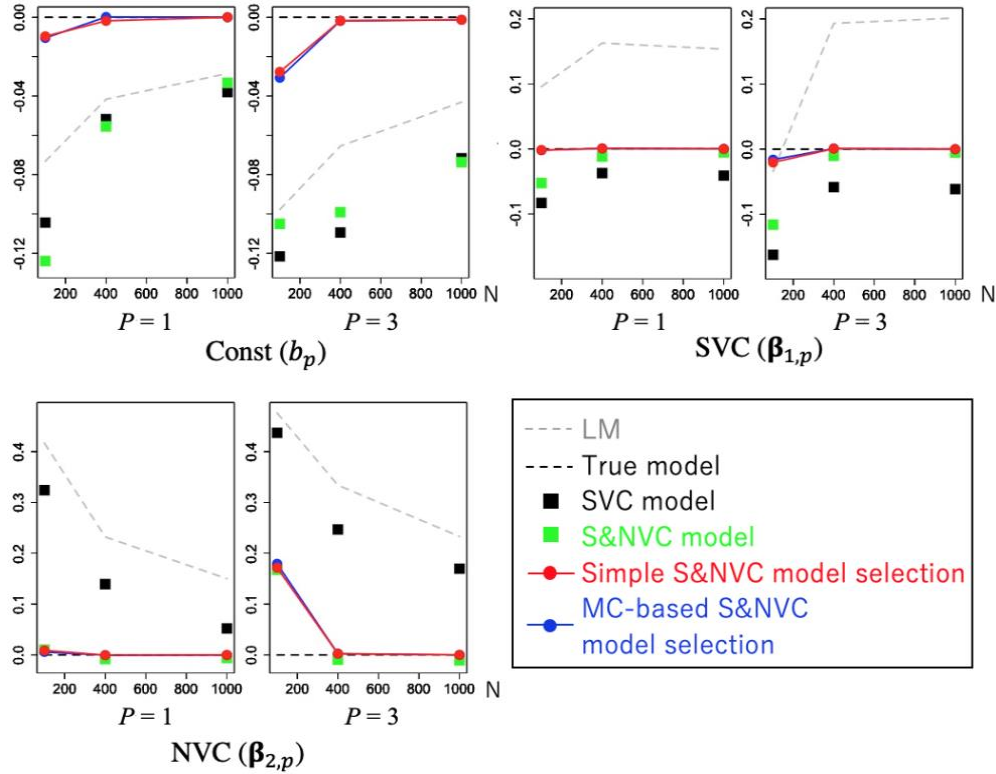

**Figure 4.** Bias of the standard error estimates

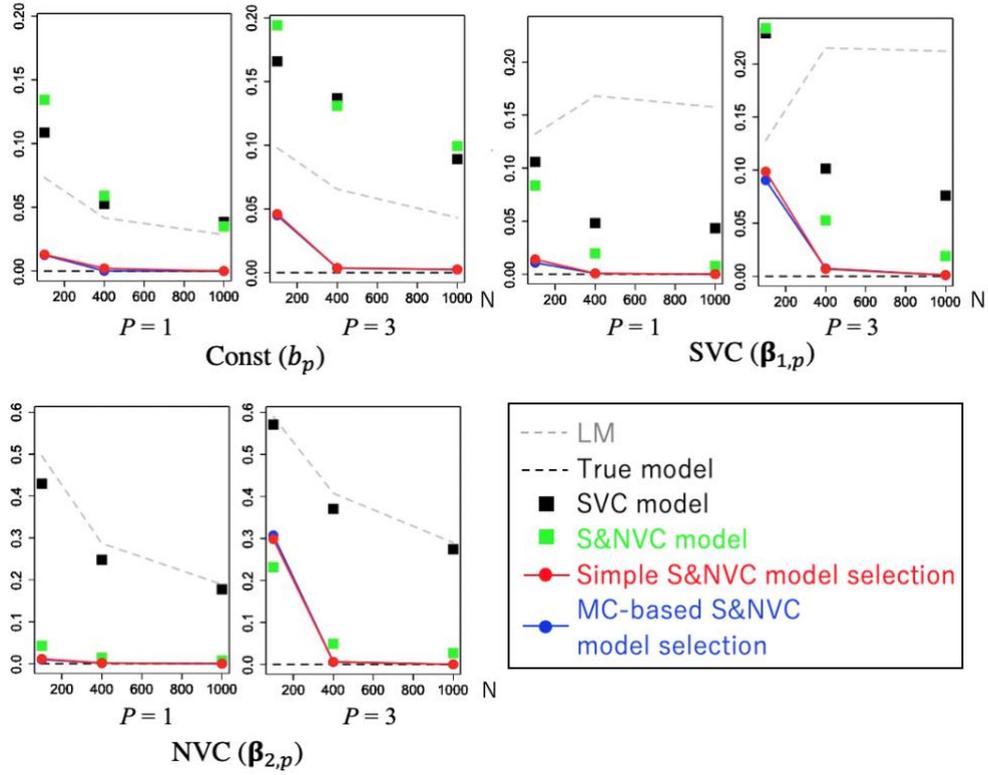

**Figure 5.** RMSE of the standard error estimates

4.3. Benchmark comparison of model selection methods

Here, we compare our model selection approach with another commonly used model selection approach. Among the existing approaches, we selected the one implemented in the mgcv package in R (https://cran.r-project.org/web/packages/mgcv/index.html) because of the following reasons: (i) mgcv is one of the most popular packages for additive mixed modeling; (ii) the effects selection procedure in mgcv is computationally highly scalable [37], and it is a sensible benchmark for testing both the accuracy and computational efficiency of our approach.

The following models are compared: LM, S&NVC model, simple S&NVC model selection, another S&NVC model estimated from mgcv (Mgcv), and double-penalty-based Mgcv model selection [37]. The double-penalty approach is the default model selection method in the mgcv package. Roughly speaking, the double-penalty approach imposes penalty parameters to select effects in addition to the usual penalty parameters in additive models determining the smoothness of each effect. [37] showed the superior accuracy of this approach over alternatives. For faster computation, we estimate Mgcv and Mgcv with the double-penalty model selection using the bam function in the mgcv package, using fast REML [38]. As in the previous section, we assume Eq. (7) as the true model and each model is iteratively fitted 200 times.

Figure 6 summaries the RMSEs for the estimated coefficients when $P = 3$ and $N \in$

{400,1,000}. For SVC and NVC, the RMSE values obtained from all the models except for LM are quite similar. Regarding the constant effect, our simple S&NVC model selection yields the lowest estimation error. Therefore, our approach is a promising alternative of Mgcv with/without the double-penalty model selection, which is now widely used.

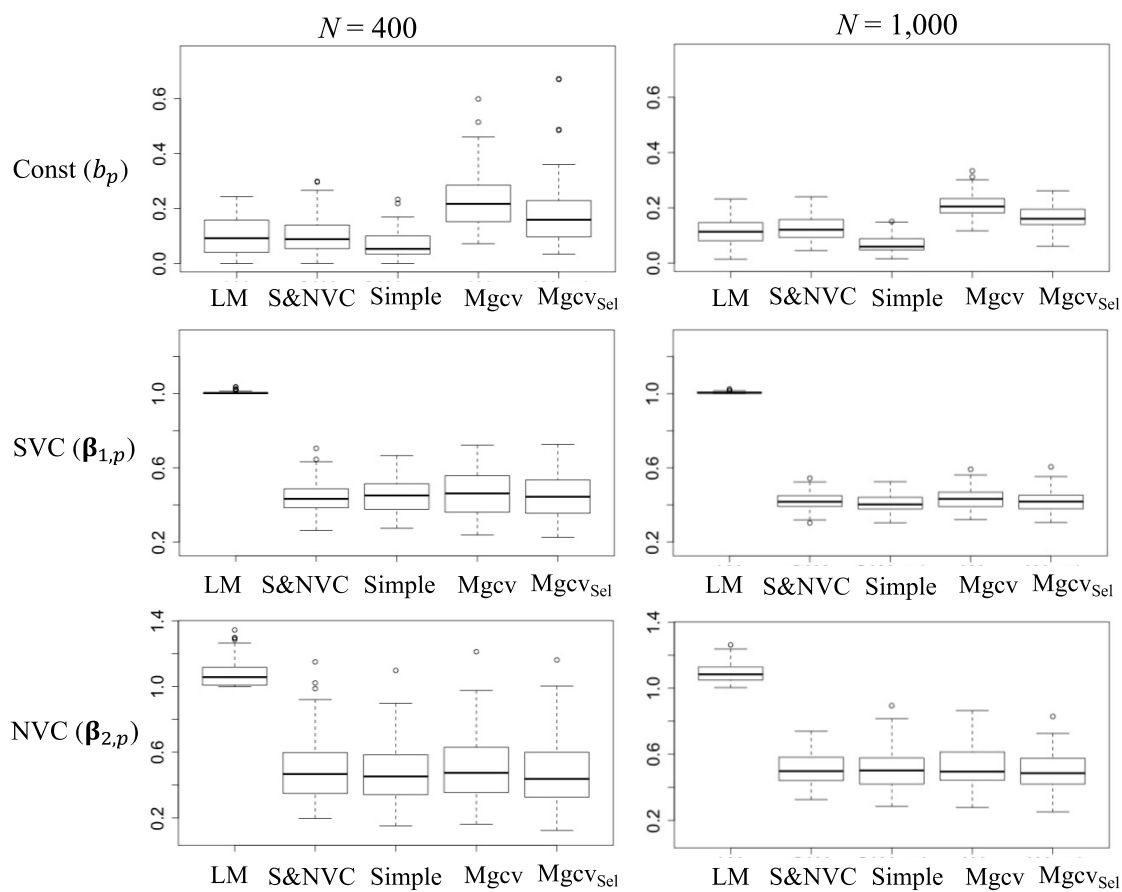

**Figure 6.** Boxplot of the RMSEs for the coefficient estimates (LM: Linear model; S&NVC: S&NVC model; Simple: Simple S&NVC model selection; Mgcv: Mgcv without model selection; Mgcv$_{sel}$: Mgcv with double-penalty-based model selection).

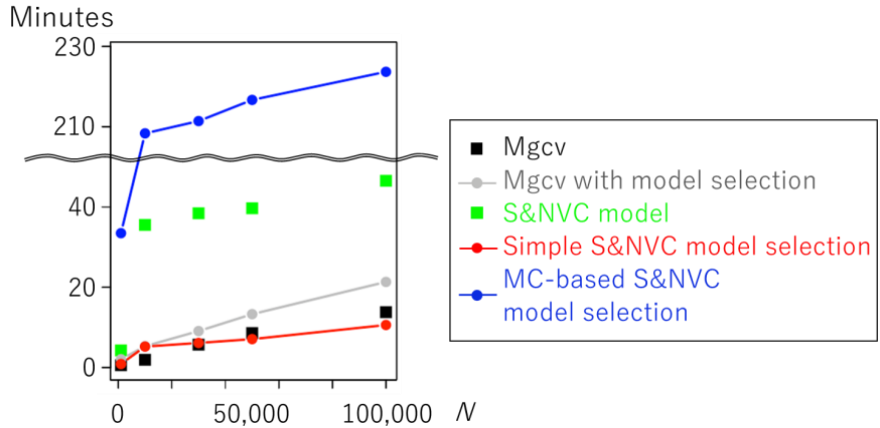

**Figure 7.** Computation time. See Figure 4 for model names. We use a Mac Pro (3.5 GHz, 6-Core Intel Xeon E5 processor with 64 GB memory). R (version 4.0.0; https://cran.r-project.org/) is used for model estimation.

Finally, Figure 7 compares the computational time in the cases of $P = 10$ and $N \in \{1,000, 10,000, 30,000, 50,000, 100,000\}$. Here, the estimations are performed five times in each case, and the resulting computation times are averaged. Mgcv with double-penalty-based model selection is slightly slower than Mgcv because the former additionally estimates the penalty parameters of the selection effects. As explained in Section 1, such additional computational time for model/effect selection is taken for granted. On the other hand, the simple S&NVC model selection approach has a considerably shorter computational time than the naïve S&NVC model because the former treats only selected effects when evaluating likelihood in each iteration, while the latter includes all the effects when evaluating likelihood. In addition, owing to the pre-conditioning procedure, the increase in computational time with respect to $N$ is suppressed with respect to the Mgcv-based alternatives. In contrast, the MC-based S&NVC model selection is slower than alternatives, as shown in Figure 7.[4] Based on estimation accuracy and computational efficiency, we recommend the simple S&NVC model selection as the default choice.

Theoretically, the simple S&NVC model selection is the fastest because of the following reasons: Monte Carlo simulation is not needed unlike the MC-based approach; smaller number of penalty parameters than Mgcv using $K$ penalty parameters for smoothness selection while another $K$ parameters for model selection. The result here confirms the computational efficiency of our simple selection approach experimentally.

In summary, the analysis results demonstrate that our approach selects effects accurately and computationally efficiently, even in comparisons with the state-of-arts approaches.

---

[4] The Monte Carlo simulation can be parallelized for fast computation although we did not do that in this comparison.

5. Application to crime modeling

5.1. Outline

This section applies a simple approach to the Dai-Tokyo Bouhan network database (https://www.bouhan.metro.tokyo.lg.jp/) provided by the Office for Promotion of Citizen Safety, Tokyo Metropolitan Government (https://www.tomin-anzen.metro.tokyo.lg.jp/english/). This database records crime statistics categorized by crime type and 1,529 minor municipal districts in Tokyo, Japan.[5] This study focuses on bicycle theft and shoplifting, the two most frequent non-burglary crimes reported between 2017 and 2018. The sample size was 12,232. Figure 8 plots the crime densities by minor municipal districts (number of incidences/km$^2$) (first quarter of 2017). The eastern area within the looping railway (Yamanote line) is the central area, and other railways extend in all directions from the looping railway, as shown in the figure. The Chuo line is one such line.

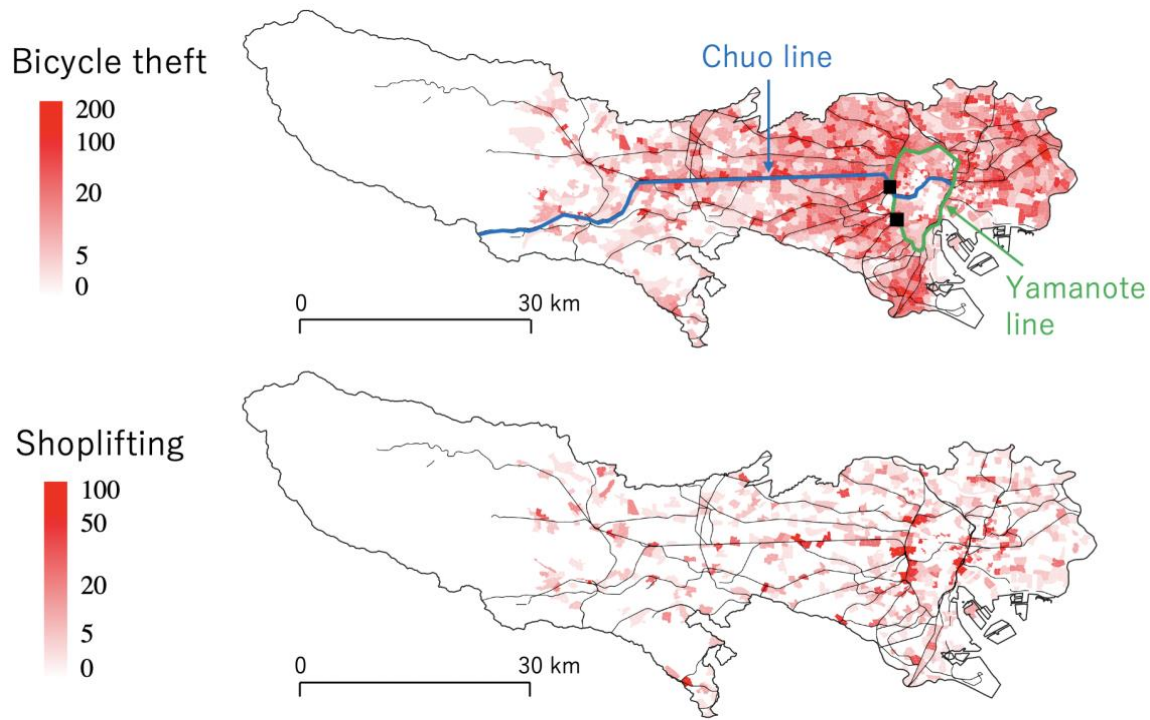

**Figure 8.** Crime density by minor municipal districts [number of incidences/km$^2$] (first quarter of 2017). Lines denote railways except the subway. The black squares in the top panel are Shinjuku station (north-west) and Tokyo station (south-east), which are major railway stations.

---

[5] The district zones are delineated to be homogeneous within each zone. Districts in urban areas are quite small while some districts in western mountain areas are large. The minimum area equals 0.003 km$^2$, the 1st quarter 0.301 km$^2$, the median 0.657 km$^2$, the 3rd quarter 1.102 km$^2$, and the maximum 68.714 km$^2$.

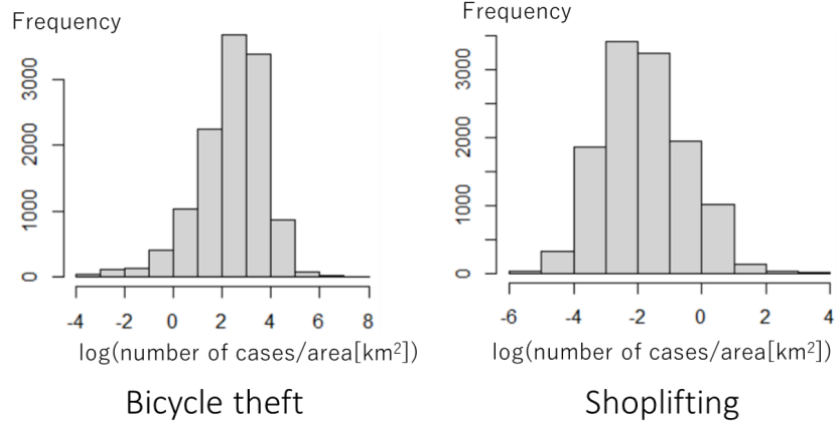

**Figure 9.** Histograms of the logged crime density per districts per quarter

We model the logged number of bicycle theft and shoplifting cases per area by districts by quarter. Based on Figure 9, which displays the histograms of the crime densities, data distributions are nearly Gaussian. A (Gaussian) linear model will be acceptable in our case. This study applies the following linear model:

$$y_{i,t}^c = \beta_{i,0} + \beta_{i,1} y_{i,t-1}^c + \beta_{i,2} Y_{i,t-1}^{-c} + \sum_{k=3}^{6} \beta_{i,k} x_{i,k} + g_{d(i)}^{(s)} + g_{q(i)}^{(t)} + \varepsilon_i, \qquad \varepsilon_i \sim N(0, \sigma^2), \qquad (10)$$

where $y_{i,t}^c$ is the number of $c$-th crime per area in the $i$-th district in the $t$-th quarter (log-scale).

We then consider explanatory variables. Based on the routine activity theory [39], the following three are the triggers for crimes: (i) potential offender, (ii) suitable target, (iii) absence of guardian. While the crime prevention level explaining (iii) is roughly equal across Tokyo, (i) and (ii) considerably changes depending on district. Regarding bicycle theft, we consider nighttime population density (Popden) as an explanatory variable approximating (i) the number of potential attackers. Since bicycle is widely used for shopping in the study area, we consider Popden and the number of retails (Retail) as explanatory variables describing (ii) the number of targets or bicycles. Regarding shoplifting, we consider daytime population density (Dpopden) as a variable explaining (i) the number of attacker while Retail as another explanatory variable explaining (ii) the number of targets.

In crime geography, near repeat victimization [40] explaining the tendency that crimes tend to repeat in similar region, is known as a common phenomenon. Such repetitive tendency occurs because of the regional heterogeneity such as resident characteristics (risk heterogeneity hypothesis) or repetitive crimes by the same group that have knowledge about the crime area (event dependency hypothesis) (see [41]). To consider such local repetitiveness, we include crime density in the previous quarter $y_{i,t-1}^c$ as an explanatory variable and call this variable Repeat. As such repetitive tendencies can occur across crime types, we also include the logged density of non-burglary crime $Y_{i,t-1}^{-c}$

(RepOther) apart from the $c$-th crime in the previous quarter. For the bicycle theft model, $Y_{i,t-1}^{-c}$ is given by log( the number of non-burglary crimes except for bicycle theft / area[km$^2$] ) while $Y_{i,t-1}^{-c}$ for the shoplifting model is similarly defined.

In addition, we include the following explanatory variables $\{x_{i,3}, x_{i,4}, x_{i,5}, x_{i,6}\}$ describing local environmental factors: ratio of foreigners among residents (Fpopden); unemployment ratio (UnEmp); ratio of residents who graduate university (Univ). These three are proxy variables of race, economic deprivation, and education. Strong influence from such local environmental factors has been demonstrated in studies relating risk terrain modeling (RTM: [42]). These data were collected from the National Census Statistics by Minor Municipal Districts in 2015.

As in the previous section, we assume a spatially varying intercept $\beta_{i,0}$ to eliminate residual spatial dependence and select coefficient type $\{\beta_{i,1}, \dots, \beta_{i,6}\}$ among {constant, SVC, NVC, S&NVC} using a simple approach. Furthermore, to capture the heterogeneity among individual districts and time periods, we consider $g_{d(i)}^{(s)} \sim N(0, \tau_{(s)}^2)$ and $g_{q(i)}^{(t)} \sim N(0, \tau_{(t)}^2)$, respectively, where $\tau_{(s)}^2$ and $\tau_{(t)}^2$ are variance parameters. These terms represent the group effects of district $d(i)$ and quarter $q(i)$ in which the $i$-th sample is observed. They are intercepts by districts and quarter, respectively (e.g., $g_{d(i)}^{(s)}$ takes larger positive values in districts with larger number of crimes). The inclusion or exclusion of $\{g_{d(i)}^{(s)}, g_{q(i)}^{(t)}\}$ is also automatically selected by the simple approach.

5.2. Coefficient estimation results

The (conditional) adjusted R-squared value is 0.914 for the bicycle theft model and 0.928 for the shoplifting model. The accuracy of these models was verified. The estimation of these models took 67.3 seconds and 52.0 seconds, respectively, confirming the computational efficiency of our approach.

Tables 2 and 3 summarize the estimated coefficients and their statistical significance. For bicycle theft, the variables Repeat, RepOther, and Popden are positively statistically significant at the 1 % level across districts. The result suggests the strong tendency of the near repeat victimization not only within the same crime type but also across crime types. Given that a higher population density implies larger number of potential offenders and targets (see Section 5.1), the positive sign of Popden is intuitively reasonable. The selected coefficient type for Repeat is S&NVC, while those for RepOther and Popden are NVCs. Although NVC has rarely been considered in spatial modeling, this result indicates the importance of considering NVC. Fpopden, UnEmp, and Univ, whose coefficients are estimated to be constant, are statistically insignificant (Table 3). Among the group effects, $g_{q(i)}^{(t)}$ is selected. In short, near repeat victimization, population, and season are the dominant determinants whereas the number of foreign people, economic status, and education are not. Later, we will look the estimated coefficients in more detail.

Table 2. Summary of Coefficient Estimates

| | Bicycle theft | | | | | | | |
|---|---|---|---|---|---|---|---|---|
| Coefficients | Intercept | Repeat | RepOther | Popden | Retail | Fpopden | UnEmp | Univ |
| Minimum | -0.469 | 0.702 | 0.144 | 0.008 | | | | |
| 1st quantile | -0.387 | 0.765 | 0.160 | 0.017 | | | | |
| Median | -0.313 | 0.787 | 0.188 | 0.021 | 0.014 | -0.015 | 0.028 | -0.035 |
| 3rd quantile | -0.283 | 0.805 | 0.208 | 0.025 | | | | |
| Maximum | -0.208 | 0.872 | 0.279 | 0.031 | | | | |
| | Shoplifting | | | | | | | |
| Coefficients | Intercept | Repeat | RepOther | Dpopden | Retail | Fpopden | UnEmp | Univ |
| Minimum | -0.312 | 0.854 | | | | | | |
| 1st quantile | -0.295 | 0.886 | | | | | | |
| Median | -0.286 | 0.894 | 0.125 | $8.18 \times 10^{-5}$ | $5.83 \times 10^{-4}$ | -0.027 | 0.135 | 0.035 |
| 3rd quantile | -0.261 | 0.903 | | | | | | |
| Maximum | -0.226 | 0.934 | | | | | | |

Table 3. Proportion of statistical significance levels in each of the coefficients

| | Bicycle theft | | | | | | | |
|---|---|---|---|---|---|---|---|---|
| Significance | Intercept | Repeat | RepOther | Popden | Retail | Fpopden | UnEmp | Univ |
| 10% level | 0.000 | 0.000 | 0.000 | 0.000 | | | | |
| 5% level | 0.000 | 0.000 | 0.000 | 0.000 | 0.000 | 0.000 | 0.000 | 0.000 |
| 1% level | 1.000 | 1.000 | 1.000 | 1.000 | | | | |
| | Shoplifting | | | | | | | |
| Significance | Intercept | Repeat | RepOther | Dpopden | Retail | Fpopden | UnEmp | Univ |
| 10% level | 0.000 | 0.000 | 0.000 | 0.000 | | | | |
| 5% level | 0.000 | 0.000 | 0.000 | 0.000 | 0.000 | 0.000 | 0.000 | 0.000 |
| 1% level | 1.000 | 1.000 | 1.000 | 1.000 | | | | |

For shoplifting, Repeat and RepOther are again positively significant, suggesting that not only shoplifting but also other non-burglary crimes in a quarter increases shoplifting in the next quarter. Dpopden is also positively significant. This suggests that shoplifting increases in central areas where people concentrate during daytime. Retail, Fpopden, UnEnp, Univ, and the two group effects are insignificant. For the significant variables, the selected type of coefficients are as follows: S&NVC for Repeat; NVC for RepOther and Dpopden; and constant for the others. In sum, repeat tendency and daytime population are shown to be significant determinants for shoplifting.

Figure 10 plots the estimated S&NVC on Repeat. The coefficients for bicycle theft decrease in the central area and southwest of the center. These areas are affluent areas. Bicycle theft is expected to be less repeated in these areas. Conversely, in the middle area, the coefficients increase along the Chuo line, which is a major railway (Figure 8) line. The repetitive tendency is especially strong near major stations on the line. Because subways and bus routes are densely spread in the affluent area (although not shown in the figure), bicycle user is relatively limited in this area. In addition, most residents in this area are affordable to buy bicycle. These properties might lower the number of potential offenders for bicycle theft. By contrast, in the middle area, railways and bus routes are relatively limited, and many people use bicycle for shopping or going to railway station; bicycle is really needed in this area. In addition, the income level in this area is low relative to the affluent area. These properties might increase potential offenders.

As for shoplifting, the coefficients on Repeat increase in the central area where the Yamanote line runs (see Figure 8). This area includes central commercial districts near the Shinjuku and Tokyo stations. A wide variety of products are sold in these districts. Besides, luxury stores are concentrated in these areas. Such properties might attract potential offenders and increase near repeat victimization.

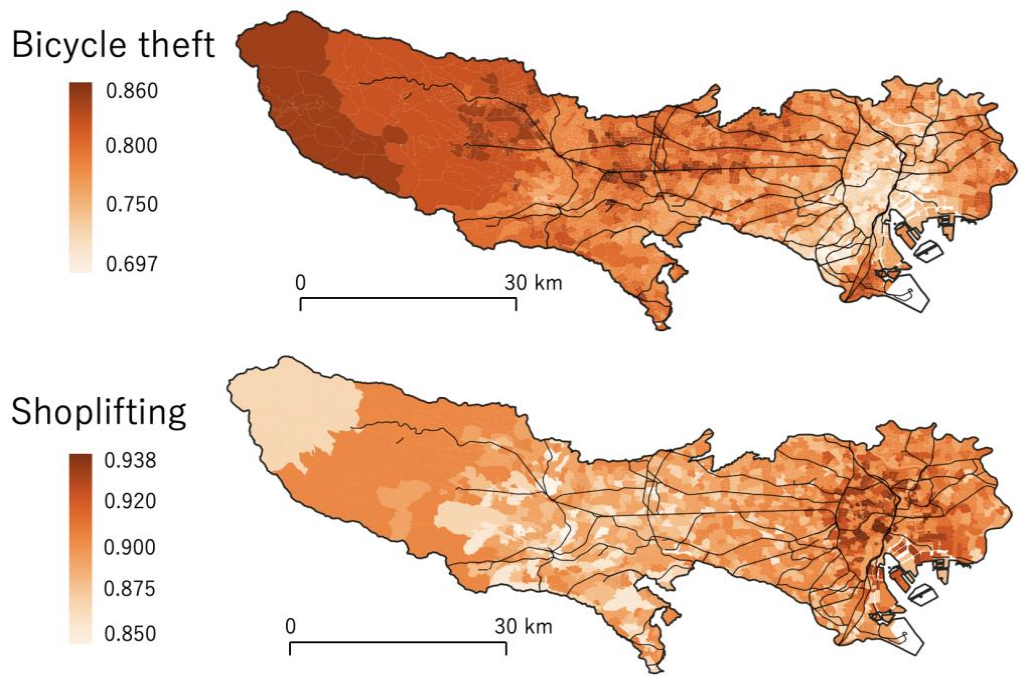

**Figure 10.** Estimated S&NVC on Repeat

Figure 11 displays the estimated NVCs. For both bicycle theft and shoplifting, the NVC on Repeat, which is in S&NVC, takes a high value if Repeat is high (see the left two panels of Figure 11). This means that repetitive tendency becomes strong if there were many crimes in the previous quarter. In contrast, smaller RepOther has higher coefficients. This means that the number of bicycle thefts and shoplifting are correlated with other crimes in low-risk areas (i.e., low RepOther values). For bicycle theft, higher Popden and Dpopden have lower coefficients. This means that bicycle thefts increase as population grows, but the rate of increase declines as population increases.

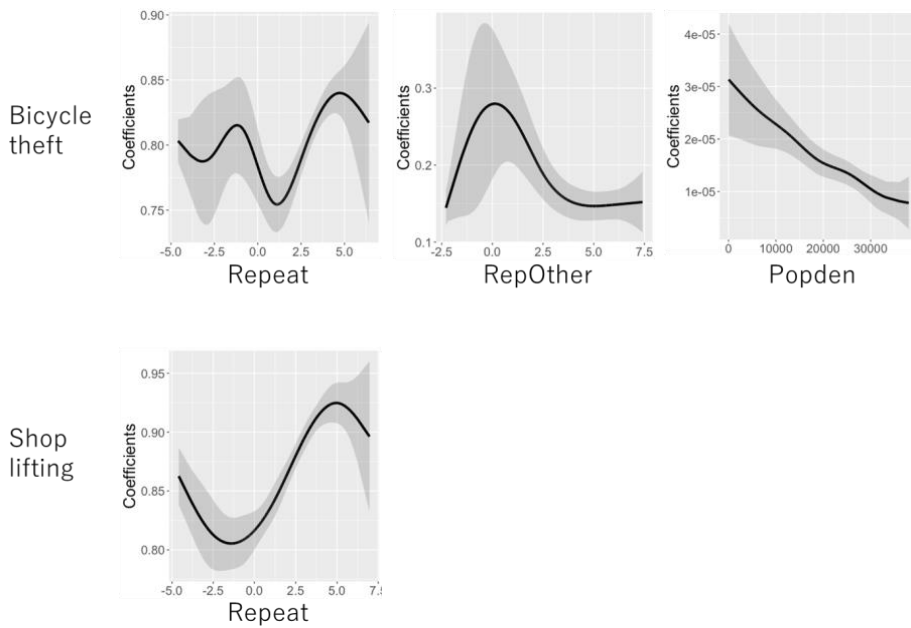

**Figure 11.** Estimated NVCs. Solid lines represent coefficient estimates and the grey areas represent 95 % confidence intervals. For Repeat, NVC is extracted from S&NVC.

Table 4. Estimated group effects

|  | Bicycle theft | | |
| --- | --- | --- | --- |
|  | Estimate | Standard Error | t-value |
| Jan-Mar | -0.104 | 0.058 | -1.812 |
| Apr-Jun | 0.075 | 0.058 | 1.292 |
| Jul-Sep | 0.028 | 0.057 | 0.493 |
| Oct-Dec | 0.001 | NA | NA |

Finally, Table 4 summarizes the estimated group effects on bicycle thefts by quarter. The results on shoplifting are not shown because both group effects were not selected. One interesting finding is that bicycle theft cases decrease in the first quarter (January to March), while it increases in the second quarter (April to June). This suggests that bicycle theft cases increase in the second quarter (April–June) probably because of the warm climate increasing bicycle user and/or change in lifestyle and routine actives in the spring season (see [43]).

While spatial regression models have been used for crime modeling, their results strongly depend on model assumptions. For example, studies using GWR always discuss coefficient estimates smoothly varying over geographical space while spatial econometric studies discuss constant coefficient estimates in the presence of spatial dependence. In contrast, our result, which selected coefficients types, is less dependent on such modeling assumptions. Our result is more reliable. In addition, our result demonstrating the strong impact from NVC, which has been overlooked in spatial statistics, rather than SVC, suggests the dangerousness of using a spatial regression model without model selection. Our model selection approach will be useful to increase reliability of spatial regression analysis.

5.3. Application to crime prediction

Crime prediction for the near future is important for preventing crimes. Here, we apply our model ($SNM_{Sel}$) to predict the number of bicycle theft and shoplifting cases by minor municipal districts in the first quarter of 2019 by employing the model trained with the data between 2017 and 2018. The accuracy is compared with the kernel density estimation (KDE), which is a popular crime prediction method in this area [44]. The ks package (https://cran.r-project.org/web/packages/ks/index.html) in R was used for the KDE model estimation and prediction. The plug-in selector of [45] is used to optimize the kernel bandwidth.

Figures 12 and 13 compare the predicted and actual number of bicycle theft and shoplifting cases per area, respectively. For both cases, the KDE results are oversmoothed. The RMSE values are 5.77 and 6.93, respectively. Conversely, owing to the inclusion of spatial and non-spatial effects, our approach appropriately detects local hot spots. The resulting map pattern is quite similar to the observed numbers. The RMSEs are 3.53 for bicycle theft and 1.62 for shoplifting, which are considerably lower than those of KDE. This suggests that our approach is useful for crime prevention, such as the design of efficient patrol routes.

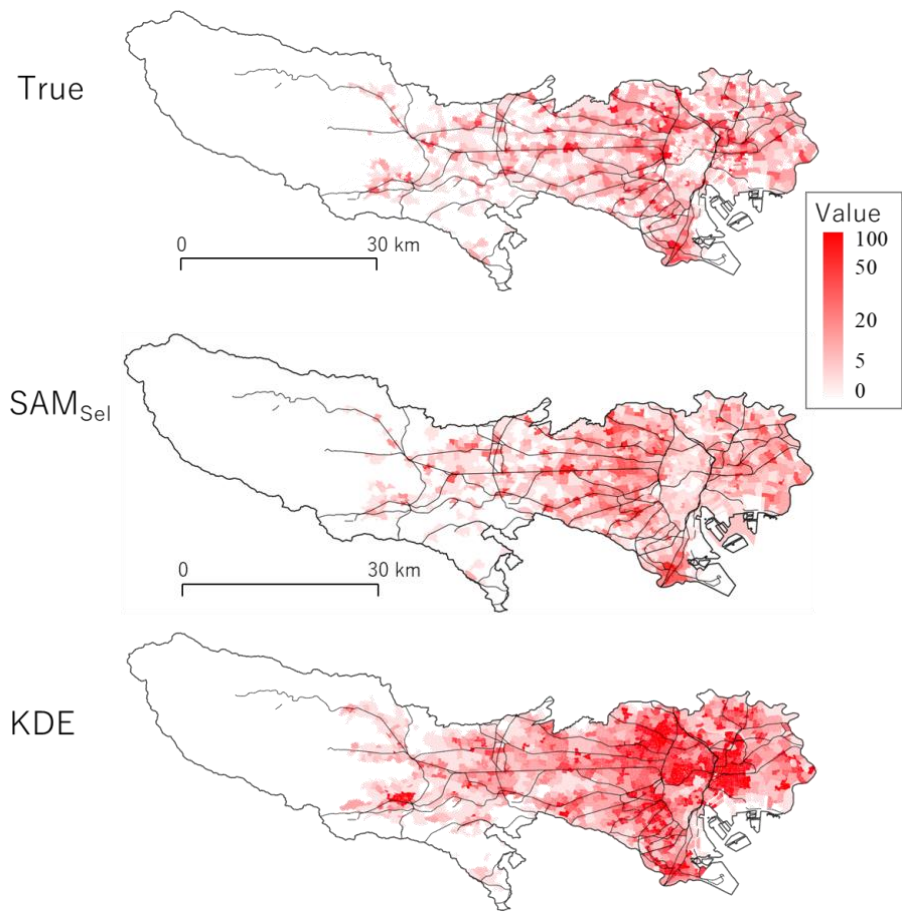

**Figure 12.** Number of bicycle theft cases in the first quarter of 2019 and the predicted results.

In short, we have demonstrated the usefulness or our approach taking crime modeling as an example. While we considered data noise by introducing a noise term, we could not consider potential data bias attributable to under-recording/reporting, survey questions for crime survey whose minor change can dramatically change answers, and other reasons (see, e.g., [46,47]). Consideration of bias in crime data will be an important next step for a more reliable crime modeling. In addition, we considered the repetitiveness at a district level. However, based on studies on repeat victimization [48,49], such repetitiveness is more prominent in a finer spatial scale such as individual property, street segment, and 250-meter grid. It is another important future task to consider more disaggregated spatial crime data rather than district-level data. Given that PredPol (https://www.predpol.com/) and many other crime prediction systems operate in a finer spatial scale like 250 m grids, extension of our crime model for spatially finer prediction will also be a remaining task toward social implementation.

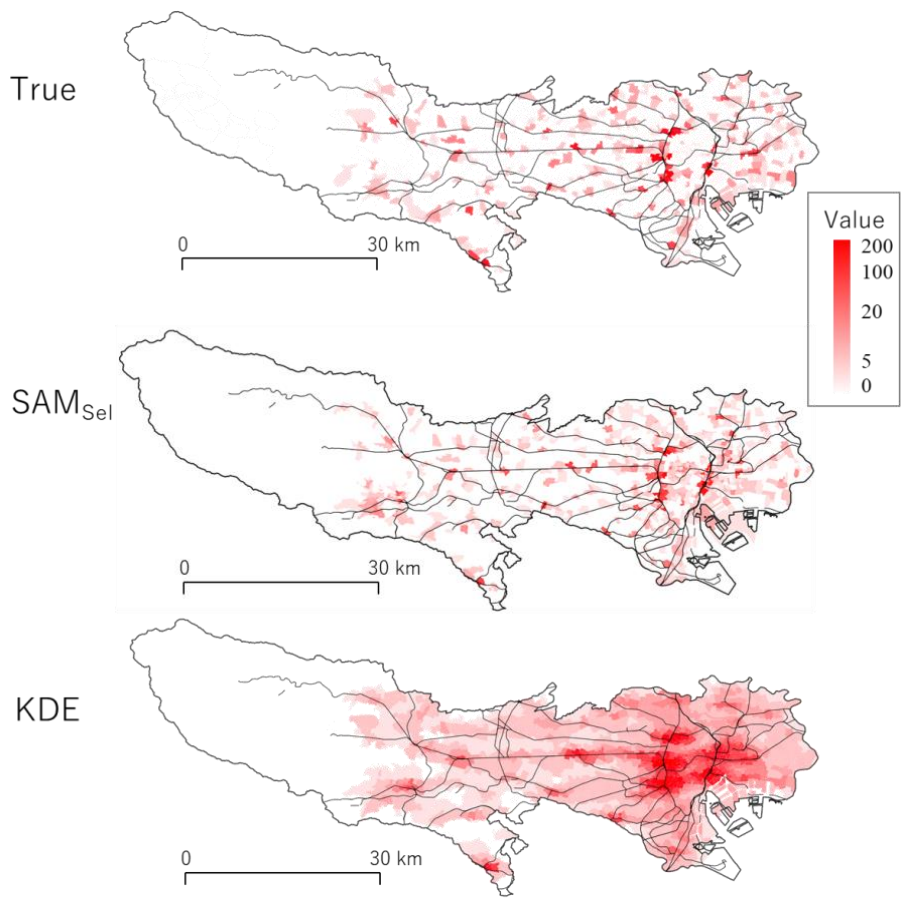

**Figure 13.** Number of shoplifting cases in the first quarter of 2019 and the predicted results.

## 6. Concluding remarks

This study develops a model or coefficient type selection approach for spatial additive mixed models. The simulation experiments suggest that the present simple approach accurately selects the true model. Moreover, even though model selections usually increase computational time, our model selection dramatically reduces it. This property will be valuable for estimating a complex model that involves many candidate effects from large samples. The criminal analysis demonstrates that our approach provides intuitively reasonable results. Our approach highlights and quantifies numerous hidden effects behind geographic phenomena. These prominent features will be useful for a wide spectrum of analyses, such as crime analysis, ecological studies, and environmental studies.

However, there are many issues to be addressed. First, we need to incorporate spatio-temporal variations in regression coefficients or residuals. Spatially and temporally varying coefficients (STVC) have been studied by [50-52], among others. However, the discussion on selecting SVC, STVC, or other coefficients is still limited in spatial statistics. The spatio-temporal process, which comprises pure-spatial, pure-temporal, and spatio-temporal interaction processes, is much more complex than purely spatial variation. A well-manipulated selection of SVC, STVC, or other coefficient types will be very important in improving the accuracy and stability of spatio-temporal modeling. The consideration of dynamic temporal behavior is important therein [53,54]. Second, it is important to consider a larger number of coefficients, such as over 100. Sparse modeling approaches, such as LASSO [55] and smoothly cropped absolute deviation (SCAD) penalty [56] will be helpful in estimating SVC, NVC, or other varying coefficients while avoiding overfitting. Third, we need to extend our approach for non-Gaussian data modeling while maintaining the computational efficiency. For example, extension for Poisson or negative binomial regression will be an important future task to model crime count rather than crime density (see e.g. [1]) while regression with skewed distribution will be useful for modeling maximum temperature or other data relating extreme value (see e.g. [57]).

The developed model selection procedure was implemented in an R package spmoran (https://cran.r-project.org/web/packages/spmoran/index.html).

Acknowledgement

These research results were obtained from the research commissioned by the National Institute of Information and Communications Technology (NICT), JAPAN.

Reference

1. Osgood, D.W. Poisson-based regression analysis of aggregate crime rates. *Journal of quantitative criminology* **2000**, *16*, 21–43.
2. Cahill, M.; Mulligan, G. Using geographically weighted regression to explore local crime patterns. *Social Science Computer Review* **2007**, *25*, 174–193.
3. Bernasco, W.; Block, R. Robberies in Chicago: A block-level analysis of the influence of crime generators, crime attractors, and offender anchor points. *Journal of Research in Crime and Delinquency* **2011**, *48*, 33–57.
4. Maguire, M.; McVie, S. Crime data and criminal statistics: A critical reflection. In *The Oxford Handbook of Criminology*; Maruna, S., McAra, L., Eds.; Oxford University Press, UK, **2017**; pp. 163–89.
5. LeSage, J.P.; Pace R.K. *Introduction to Spatial Econometrics*; CRC Press: Boca Raton, US, 2009.
6. Cressie, N.; Wikle, C.K. *Statistics for Spatio-Temporal Data*; John Wiley & Sons: Hoboken, Canada, 2011.
7. Brunsdon, C.; Fotheringham, S.; Charlton, M. Geographically weighted regression. *Journal of the Royal Statistical Society: Series D (The Statistician)* **1998**, *47*, 431–443.
8. Fotheringham, A.S.; Brunsdon, C.; Charlton, M. *Geographically Weighted Regression: The*

*Analysis of Spatially Varying Relationship*; John Wiley & Sons: West Sussex, UK, 2002.

9. Lee, S.; Kang, D.; Kim, M. Determinants of crime incidence in Korea: a mixed GWR approach. *Proceedings of the World Conference of the Spatial Econometrics Association* **2009**, 8–10.

10. Arnio, A.N.; Baumer, E.P. Demography, foreclosure, and crime: Assessing spatial heterogeneity in contemporary models of neighborhood crime rates. *Demographic Research* **2012**, *26*, 449–486.

11. Umlauf, N.; Adler, D.; Kneib, T.; Lang, S.; Zeileis, A. Structured additive regression models: An R interface to BayesX. *Journal of Statistical Software* **2015**, *21*, 63.

12. Nakaya, T.; Fotheringham, S.; Charlton, M.; Brunsdon, C. Semiparametric geographically weighted generalised linear modelling in GWR 4.0. *Proceeding of Geocomputation* **2009**.

13. Wheeler, D.C. Simultaneous coefficient penalization and model selection in geographically weighted regression: the geographically weighted lasso. *Environment and planning A* **2009**, *41*, 722–742.

14. Comber, A.; Brunsdon, C.; Charlton, M.; Dong, G.; Harris, R.; Lu, B.; Lu, Y.; Murakami, D.; Nakaya, T.; Wang, Y.; Harris, P. The GWR route map: a guide to the informed application of Geographically Weighted Regression. *ArXiv* **2020**, 2004.06070.

15. Huang, J.; Horowitz, J.L.; Wei, F. Variable selection in nonparametric additive models. *Annals of statistics* **2010**, *38*, 2282.

16. Amato, U.; Antoniadis, A.; De Feis, I. Additive model selection. *Statistical Methods & Applications* **2016**, *25*, 519–654.

17. Mei, C.L.; He, S.Y.; Fang, K.T. A note on the mixed geographically weighted regression model. *Journal of Regional Science* **2004**, *44*, 143–157.

18. Li, Z.; Fotheringham, A.S.; Li, W.; Oshan, T. Fast Geographically Weighted Regression (FastGWR): a scalable algorithm to investigate spatial process heterogeneity in millions of observations. *International Journal of Geographical Information Science* **2019**, *33*, 155–175.

19. Murakami, D.; Griffith, D.A. Spatially varying coefficient modeling for large datasets: Eliminating N from spatial regressions. *Spatial Statistics* **2019**, *30*, 39–64.

20. Murakami, D.; Griffith, D.A. A memory-free spatial additive mixed modeling for big spatial data. *Japanese Journal of Statistics and Data Science* **2020**, 1–27.

21. Murakami, D.; Yoshida, T.; Seya, H.; Griffith, D. A.; Yamagata, Y. A Moran coefficient-based mixed effects approach to investigate spatially varying relationships. *Spatial Statistics* **2017**, *19*, 68–89.

22. Tiefelsdorf, M., & Griffith, D. A. (2007). Semiparametric filtering of spatial autocorrelation: the eigenvector approach. *Environment and Planning A*, *39*(5), 1193-1221.

23. Griffith, D.A. *Spatial Autocorrelation and Spatial Filtering: Gaining Understanding Through Theory and Scientific Visualization*; Springer Science & Business Media: Berlin, Germany, 2003.

24. Murakami, D.; Griffith, D.A. Balancing spatial and non-spatial variation in varying coefficient modeling: a remedy for spurious correlation. *ArXiv* **2020**, 2005.09981.
25. Wheeler, D.; Tiefelsdorf, M. Multicollinearity and correlation among local regression coefficients in geographically weighted regression. *Journal of Geographical Systems* **2005**, *7*, 161–187.
26. Bates, D.; Mächler, M.; Bolker, B.; Walker, S. Fitting linear mixed-effects models using lme4. *ArXiv* **2014**, 1406.5823.
27. Winter, B.; Wieling, M. How to analyze linguistic change using mixed models. *Growth Curve Analysis and Generalized Additive Modeling. Journal of Language Evolution* **2016**, *1*, 7–18.
28. Baayen, H.; Vasishth, S.; Kliegl, R.; Bates, D. The cave of shadows: Addressing the human factor with generalized additive mixed models. *Journal of Memory and Language* **2017**, *94*, 206–234.
29. Gurka, M.J. Selecting the best linear mixed model under REML. *The American Statistician* **2006**, *60*, 19–26.
30. Müller, S.; Scealy, J.L.; Welsh, A.H. Model selection in linear mixed models. *Statistical Science* **2013**, *28*, 135–167.
31. Dimova, R.B.; Markatou, M.; Talal, A.H. Information methods for model selection in linear mixed effects models with application to HCV data. *Computational statistics & data analysis* 2011, *55*, 2677–2697.
32. Sakamoto, W. Bias-reduced marginal Akaike information criteria based on a Monte Carlo method for linear mixed-effects models. *Scandinavian Journal of Statistics* **2019**, *46*, 87–115.
33. Greven, S.; Kneib, T. On the behaviour of marginal and conditional AIC in linear mixed models. *Biometrika* **2010**, *97*, 773–789.
34. Belitz, C.; Lang, S. Simultaneous selection of variables and smoothing parameters in structured additive regression models. *Computational Statistics & Data Analysis* **2008**, *53*, 61–81.
35. Reiss, P.T.; Todd Ogden, R. Smoothing parameter selection for a class of semiparametric linear models. *Journal of the Royal Statistical Society: Series B (Statistical Methodology)* **2009**, *71*, 505–523.
36. Wood, S.N. Fast stable restricted maximum likelihood and marginal likelihood estimation of semiparametric generalized linear models. *Journal of the Royal Statistical Society: Series B (Statistical Methodology)* **2011**, *73*, 3–36.
37. Marra, G.; Wood, S.N. Practical variable selection for generalized additive models. *Computational Statistics & Data Analysis* **2011**, *55*, 2372–2387.
38. Wood, S.N.; Li, Z.; Shaddick, G.; Augustin, N.H. Generalized additive models for gigadata: modeling the UK black smoke network daily data. *Journal of the American Statistical Association* **2017**, *112*, 1199–1210.
39. Felson, M.. *Crime and Everyday Life: Insights and Implications for Society (The Pine Forge*

*Press Social Science Library)* **1994**, Pine Forge.

40. Farrell, G. Preventing repeat victimization. *Crime and Justice* **1995**, *19*, 469–534.
41. Johnson, S.D. Repeat burglary victimisation: a tale of two theories. *Journal of Experimental Criminology* **2008**, *4*, 215–240.
42. Caplan, J.M., Kennedy, L.W.; Miller, J. Risk terrain modeling: Brokering criminological theory and GIS methods for crime forecasting. *Justice quarterly* **2011**, *28*, 360–381.
43. Ranson, M. Crime, weather, and climate change. *Journal of Environmental Economics and Management* **2014**, *67*, 274–302.
44. Harada, Y.; Shimada, T. Examining the impact of the precision of address geocoding on estimated density of crime locations. *Computers & Geosciences* **2006**, *32*, 1096–1107.
45. Wand, M.P.; Jones, M.C. Multivariate plug-in bandwidth selection. *Computational Statistics* **1994**, *9*, 97–116.
46. Yu, O.; Zhang, L. The under-recording of crime by police in China: a case study. *Policing: An International Journal* 1999, *22*, 252–264.
47. Tabarrok, A.; Heaton, P.; Helland, E. The measure of vice and sin: A review of the uses, limitations, and implications of crime data. *Handbook on the Economics of Crime* **2010**, *3*, 53–81.
48. Farrell, G.; Phillips, C.; Pease, K. Like taking candy-why does repeat victimization occur. *British Journal of Criminology* **1995**, *35*, 384–399,
49. Farrell, G.; Pease, K. *Repeat Victimization*. Criminal Justice Press, NY, **2001**.
50. Gelfand, A.E.; Kim, H.J.; Sirmans, C.F.; Banerjee, S. Spatial modeling with spatially varying coefficient processes. *Journal of the American Statistical Association* **2003**, *98*, 387–396.
51. Huang, B.; Wu, B.; Barry, M. Geographically and temporally weighted regression for modeling spatio-temporal variation in house prices. *International Journal of Geographical Information Science* **2010**, *24*, 383–401.
52. Fotheringham, A.S.; Crespo, R.; Yao, J. Geographical and temporal weighted regression (GTWR). *Geographical Analysis* **2015**, *47*, 431–452.
53. Mohler, G. Modeling and estimation of multi-source clustering in crime and security data. *The Annals of Applied Statistics* **2013**, *7*, 1525–1539.
54. Kajita, M.; Kajita, S. Crime prediction by data-driven Green's function method. *International Journal of Forecasting* **2020**, *36*, 480–488.
55. Hastie, T.; Tibshirani, R.; Wainwright, M. *Statistical Learning with Sparsity: The Lasso and Generalizations*. CRC Press: New York, 2015.
56. Fan, J.; Li, R. Variable selection via nonconcave penalized likelihood and its oracle properties. *Journal of the American statistical Association* **2001**, *96*, 1348–1360.
57. Cooley, D. Extreme value analysis and the study of climate change. *Climatic change* **2009**, *97*,

(1-2), 77.

## Appendix A. Restricted log likelihood function of the spatial additive mixed model

The fast REML maximizes the marginal likelihood $loglik_R(\Theta)$ defined by integrating $\{\mathbf{b}, \mathbf{U}\}$ from the full likelihood, which is formulated as

$$loglik_R(\Theta) = -\frac{1}{2}log\begin{vmatrix} \mathbf{X'X} & \mathbf{X'\tilde{E}(\Theta)} \\ \mathbf{\tilde{E}'(\Theta)X} & \mathbf{\tilde{E}'(\Theta)\tilde{E}(\Theta)+I} \end{vmatrix} - \frac{N-K}{2}\left(1+log\left(\frac{2\pi d(\Theta)}{N-K}\right)\right), \quad (A1)$$

where $\mathbf{\tilde{E}}(\Theta) = [\mathbf{x}_1 \circ \mathbf{E}_1 \mathbf{V}_1(\theta_1), \ldots, \mathbf{x}_P \circ \mathbf{E}_P \mathbf{V}_P(\theta_P)]$. In the second term of Eq. (6), $d(\Theta) = \hat{\mathbf{\varepsilon}}'\hat{\mathbf{\varepsilon}} + \sum_{p=1}^{P}\hat{\mathbf{u}}'_p\hat{\mathbf{u}}_p$ balances noise variance and random effects variance, where

$$\hat{\mathbf{\varepsilon}} = \mathbf{y} - \mathbf{X}\hat{\mathbf{b}} - \mathbf{\tilde{E}}(\Theta)\hat{\mathbf{U}} \quad (A2)$$

$$\begin{bmatrix}\hat{\mathbf{b}}\\\hat{\mathbf{U}}\end{bmatrix} = \begin{bmatrix} \mathbf{X'X} & \mathbf{X'\tilde{E}(\Theta)} \\ \mathbf{\tilde{E}'(\Theta)X} & \mathbf{\tilde{E}'(\Theta)\tilde{E}(\Theta)+I} \end{bmatrix}^{-1}\begin{bmatrix}\mathbf{X'y}\\\mathbf{\tilde{E}'(\Theta)y}\end{bmatrix} \quad (A3)$$

The additive mixed model is readily estimated by first: (i) estimating $\hat{\Theta}$ by maximizing $loglik_R(\Theta)$, and then, (ii) estimating the fixed and random coefficients $[\hat{\mathbf{b}}', \hat{\mathbf{U}}']'$ by substituting $\hat{\Theta}$ into Eq.(8). One major difficulty is the computational cost of estimating $\hat{\Theta}$ in step (i) because the cost increases exponentially with respect to the number of parameters in $\hat{\Theta}$. The computational time can be disappointingly long, even for 10 variance parameters. Unfortunately, to flexibly capture the influence of many covariates, 10 or more variance parameters are typically required.

## Appendix B. Details of the model selection approaches

This appendix explains the computational details of the simple model selection method. Then, the details of the MC method are explained.

In step (a) of the simple method (see Section 3.2.1), the data matrices $\{\mathbf{y}, \mathbf{X}, \mathbf{E}_1, \ldots, \mathbf{E}_P\}$ are replaced with the following inner products: $\mathbf{M}_{0,0} = \mathbf{X'X}$, $\mathbf{M}_{0,p} = \mathbf{X'}(\mathbf{x}_p \circ \mathbf{E}_p)$, $\mathbf{M}_{p,\tilde{p}} = (\mathbf{x}_p \circ \mathbf{E}_p)'(\mathbf{x}_{\tilde{p}} \circ \mathbf{E}_p)$, $\mathbf{m}_0 = \mathbf{X'y}$, $\mathbf{m}_p = (\mathbf{x}_p \circ \mathbf{E}_p)'\mathbf{y}$, and $m_{y,y} = \mathbf{y'y}$, where "$\mathbf{a} \circ \mathbf{B}$" multiplies column vector $\mathbf{a}$ with each column of the $\mathbf{B}$ matrix element wise.

The restricted log-likelihood maximized in step (b) (see Section 3.2.1) is rewritten by substituting these inner products into Eq. (A1) [19]:

$$loglik_R(\Theta) = -\frac{1}{2}ln|\mathbf{P}| - \frac{N-K}{2}\left(1+ln\left(2\pi\frac{\|\hat{\mathbf{\varepsilon}}\|^2 + \sum_{p=1}^{P}\|\hat{\mathbf{u}}_p\|^2}{N-P}\right)\right), \quad (A4)$$

where

$$\mathbf{P} = \begin{bmatrix} \mathbf{M}_{0,0} & \mathbf{M}_{0,1}\mathbf{V}_1(\boldsymbol{\theta}_1) & \cdots & \mathbf{M}_{0,P}\mathbf{V}_P(\boldsymbol{\theta}_P) \\ \mathbf{V}_1(\boldsymbol{\theta}_1)\mathbf{M}_{1,0} & \mathbf{V}_1(\boldsymbol{\theta}_1)\mathbf{M}_{1,1}\mathbf{V}_1(\boldsymbol{\theta}_1) + \mathbf{I} & \cdots & \mathbf{V}_1(\boldsymbol{\theta}_1)\mathbf{M}_{1,P}\mathbf{V}_P(\boldsymbol{\theta}_P) \\ \vdots & \vdots & \ddots & \vdots \\ \mathbf{V}_P(\boldsymbol{\theta}_P)\mathbf{M}_{P,0} & \mathbf{V}_P(\boldsymbol{\theta}_P)\mathbf{M}_{P,1}\mathbf{V}_1(\boldsymbol{\theta}_1) & \cdots & \mathbf{V}_P(\boldsymbol{\theta}_P)\mathbf{M}_{P,P}\mathbf{V}_P(\boldsymbol{\theta}_P) + \mathbf{I} \end{bmatrix}, \quad (A5)$$

$$\|\hat{\boldsymbol{\varepsilon}}\|^2 = m_{y,y} - 2[\hat{\mathbf{b}}', \hat{\mathbf{u}}'_1, \cdots \hat{\mathbf{u}}'_P] \begin{bmatrix} \mathbf{m}_0 \\ \mathbf{V}_1(\boldsymbol{\theta}_1)\mathbf{m}_1 \\ \vdots \\ \mathbf{V}_P(\boldsymbol{\theta}_P)\mathbf{m}_P \end{bmatrix} + [\hat{\mathbf{b}}', \hat{\mathbf{u}}'_1, \cdots \hat{\mathbf{u}}'_P]\mathbf{P}_0 \begin{bmatrix} \hat{\mathbf{b}} \\ \hat{\mathbf{u}}_1 \\ \vdots \\ \hat{\mathbf{u}}_P \end{bmatrix}, \quad (A6)$$

$$\begin{bmatrix} \hat{\mathbf{b}} \\ \hat{\mathbf{u}}_1 \\ \vdots \\ \hat{\mathbf{u}}_P \end{bmatrix} = \mathbf{P}^{-1} \begin{bmatrix} \mathbf{m}_0 \\ \mathbf{V}_1(\boldsymbol{\theta}_1)\mathbf{m}_1 \\ \vdots \\ \mathbf{V}_P(\boldsymbol{\theta}_P)\mathbf{m}_P \end{bmatrix}. \quad (A7)$$

Eqs. (A4)–(A7) do not include any matrix whose size depends on $N$. Thus, in step (b), the likelihood is evaluated with a computational cost independent of the sample size.

Consider the $P$-th iteration of step (b) as an example. In step (b-1) of the $P$-th iteration, $loglik_R(\boldsymbol{\theta}_{P(s)}|\hat{\boldsymbol{\Theta}}_{-P(s)})$ is maximized with respect to $\boldsymbol{\theta}_{P(s)}$. Maximization involves iterative evaluation of $|\mathbf{P}|$ and $\mathbf{P}^{-1}$ whose computational complexity is equal $O\left(\left(\sum_{p=1}^P L_p\right)^3\right)$, which can be slow when considering many effects, as it is in our case. To reduce cost, Eq.(A7), including $\mathbf{P}^{-1}$ is expanded as follows:

$$\begin{bmatrix} \hat{\mathbf{b}} \\ \hat{\mathbf{u}}_1 \\ \vdots \\ \hat{\mathbf{u}}_P \end{bmatrix} = \begin{bmatrix} \tilde{\mathbf{V}}_{-P}^{-1} & \mathbf{O} \\ \mathbf{O} & \mathbf{V}_P(\boldsymbol{\theta}_{P(s)})^{-1} \end{bmatrix} \mathbf{Q}^{-1} \begin{bmatrix} \mathbf{m}_{-P} \\ \mathbf{m}_P \end{bmatrix} \\ - \begin{bmatrix} \tilde{\mathbf{V}}_{-P}^{-1}\mathbf{Q}^*_{-P,P} \\ \mathbf{V}_P(\boldsymbol{\theta}_{P(s)})^{-1}\mathbf{Q}^*_{P,P} \end{bmatrix} \left(\mathbf{V}_P(\boldsymbol{\theta}_{P(s)})^2 + \mathbf{Q}^*_{P,P}\right)^{-1}\left[\mathbf{Q}^*_{P,-P}\mathbf{m}_{-P} + \mathbf{Q}^*_{P,P}\mathbf{m}_P\right], \quad (A8)$$

where $\tilde{\mathbf{V}}_{-P} = \begin{bmatrix} \mathbf{I} & & & \\ & \mathbf{V}_1 & & \\ & & \ddots & \\ & & & \mathbf{V}_{P-1} \end{bmatrix}$ where $\mathbf{V}_p = \mathbf{V}_p(\hat{\boldsymbol{\theta}}_p)$. $\hat{\boldsymbol{\Theta}}_{-P(s)} \in \{\hat{\boldsymbol{\theta}}_1, \ldots, \hat{\boldsymbol{\theta}}_{P-1}, \hat{\boldsymbol{\theta}}_{P(n)}\}$,

which are fixed in this step is omitted for simplicity. $\mathbf{Q} = \begin{bmatrix} \tilde{\mathbf{M}}_{-P,-P} + \tilde{\mathbf{V}}_{-P}^{-2} & \tilde{\mathbf{M}}_{-P,P} \\ \tilde{\mathbf{M}}_{P,-P} & \mathbf{M}_{P,P} \end{bmatrix}$, where $\tilde{\mathbf{M}}_{-P,-P} = \begin{bmatrix} \mathbf{M}_{0,0} & \mathbf{M}_{0,1} & \cdots & \mathbf{M}_{0,P-1} \\ \mathbf{M}_{1,0} & \mathbf{M}_{1,1} + \mathbf{V}_1^{-2} & \cdots & \mathbf{M}_{1,P-1} \\ \vdots & \vdots & \ddots & \vdots \\ \mathbf{M}_{P-1,0} & \mathbf{M}_{P-1,1} & \cdots & \mathbf{M}_{P-1,P-1} + \mathbf{V}_{P-1}^{-2} \end{bmatrix}$ and $\tilde{\mathbf{M}}_{-P,P} =$

$[\mathbf{M}_{P,0} \quad \mathbf{M}_{P,1} \quad \cdots \quad \mathbf{M}_{P,P-1}]'$, and $\begin{bmatrix} \mathbf{Q}^*_{-P,-P} & \mathbf{Q}^*_{-P,P} \\ \mathbf{Q}^*_{P,-P} & \mathbf{Q}^*_{P,P} \end{bmatrix} = \mathbf{Q}^{-1}$.

On the other hand, $|\mathbf{P}|$, which is another tedious part, has the following expression:

$$|\mathbf{P}| = |\tilde{\mathbf{V}}_{-P}|^2|\mathbf{V}_p(\boldsymbol{\theta}_{P(s)})|^2|\tilde{\mathbf{M}}_{-P,-P} + \tilde{\mathbf{V}}_{-P}^{-2}||\mathbf{V}_p(\boldsymbol{\theta}_{P(s)})^{-2} + \mathbf{M}_{P,P} \\ - \tilde{\mathbf{M}}_{P,-P}(\tilde{\mathbf{M}}_{-P,-P} + \tilde{\mathbf{V}}_{-P}^{-2})^{-1}\tilde{\mathbf{M}}_{-P,P}|. \quad (A9)$$

To maximize the likelihood numerically, Eqs. (A8) and (A9) must be evaluated repeatedly while varying $\boldsymbol{\theta}_{P(s)}$. Fortunately, many elements in Eq. (A8) are unchanged even when $\boldsymbol{\theta}_{P(s)}$ is changed. As a result, if the elements that are independent of $\boldsymbol{\theta}_{P(S)}$ are evaluated *a priori*, the computational complexity for the iterative evaluation of Eq. (A8) is only $O(L_P^3)$, which is required for evaluating $\left(\mathbf{V}(\boldsymbol{\theta}_{P(S)})^2 + \mathbf{Q}_{P,P}^*\right)^{-1}$. Likewise, the complexity of the iterative evaluation of Eq.(A9) while varying $\boldsymbol{\theta}_{P(S)}$ is only $O(L_P^3)$. Thus, the model estimation step (b-1) scales extremely well for both sample size $N$ and the number of effects $P$. The same holds for NVC estimation step (b-3) too.

In the model selection step (b-2), we need to compare the cost function (e.g., BIC) of the model with the *P*-th SVC, which is estimated in step (b-1), with the model without the *P*-th SVC. For this, we also need to evaluate the likelihood of the latter model using Eq.(A4). The likelihood can be evaluated by replacing Eqs. (A8) and (A9) with Eqs. (A10) and (A11), respectively.

$$\begin{bmatrix} \hat{\mathbf{b}} \\ \hat{\mathbf{u}}_1 \\ \vdots \\ \hat{\mathbf{u}}_{P-1} \end{bmatrix} = \widetilde{\mathbf{V}}_{-P}^{-1}(\widetilde{\mathbf{M}}_{-P,-P} + \widetilde{\mathbf{V}}_{-P}^{-2})^{-1}\mathbf{m}_{-P} \tag{A10}$$

$$|\mathbf{P}| = \left|\widetilde{\mathbf{V}}_{-P}\right|^2 \left|\widetilde{\mathbf{M}}_{-P,-P} + \widetilde{\mathbf{V}}_{-P}^{-2}\right| \tag{A11}$$

All the elements in Eq.(A11) are already evaluated in step (b-1) (Eq. A6). Thus, Eq.(A11) was evaluated without any additional computational cost. Although $(\widetilde{\mathbf{M}}_{-P,-P} + \widetilde{\mathbf{V}}_{-P}^{-2})^{-1}$ must be additionally calculated to evaluate Eq. (A10), the computational complexity is $O\left(\left(\sum_{p=1}^{P-1} L_p\right)^3\right)$, which is still independent of sample size. In addition, iterative evaluation is not needed in this part because the computation cost in step (b-2) is trivial. The same holds for another model selection step (b-4).

In summary, both the estimation and model selection steps were performed computationally efficiently.